\documentclass[aps,prl,superscriptaddress]{revtex4-2}

\usepackage{hyperref}
\usepackage{textcomp}
\usepackage[autostyle]{csquotes}
\usepackage{amssymb}
\usepackage{amsthm}
\usepackage{amsfonts}
\usepackage{amssymb}
\usepackage{graphicx}
\usepackage{comment}
\usepackage{geometry}
\usepackage{algpseudocode}
\usepackage{algorithm}
\usepackage{subfiles}
\usepackage{subfig}
\usepackage{mathtools}
\usepackage{pgfplots}
\pgfplotsset{/pgf/number format/use comma,compat=newest}
\usepackage{url}
\usepackage{tikz}
\usepackage{bbm}
\usepackage{floatflt}
\usepackage[colorinlistoftodos]{todonotes}
\usetikzlibrary{arrows,decorations.pathmorphing,backgrounds,positioning,fit,petri}

\usepgfplotslibrary{fillbetween}

\begin{document}
\title{Matrix factorization with neural networks}

\author{Francesco Camilli}
\email[]{franceco.camilli2@unibo.it}
\affiliation{University of Bologna, Bologna, Italy}
\affiliation{\'Ecole normale supérieure, Paris, France}

\author{Marc Mézard}
\email[]{marc.mezard@unibocconi.it}
\affiliation{Bocconi University, Milan, Italy}

\date{\today}
\begin{abstract}
Matrix factorization is an important mathematical problem encountered in the context of dictionary learning, recommendation systems and machine learning. We introduce a new `decimation' scheme that maps it to neural network models of associative memory and provide a detailed theoretical analysis of its performance, showing that decimation is able to factorize extensive-rank matrices and to denoise them efficiently. We introduce a decimation algorithm based on ground-state search of the neural network, which shows performances that match the theoretical prediction.
\end{abstract}


\maketitle
Matrix factorization is a generic inference problem that is simply stated. Consider a matrix $\mathbf{Y}$ which is a  noisy observation  of the product of two matrices $\mathbf{Y}^*=\mathbf{A}\mathbf{B}$, and some prior knowledge on the distribution of elements of the two factors $\mathbf{A}$ and $\mathbf{B}$. One would like to answer two main questions : \emph{(i)} in what regimes of sizes of $\mathbf{A}$, $\mathbf{B}$ and noise is it possible to reconstruct the two factors (up to a permutation of the lines of $\mathbf{A}$ and the columns of $\mathbf{B}$) ? \emph{(ii)} Do there exist efficient algorithms that achieve a good performance?

Matrix factorization is important for several concrete tasks. In \emph{dictionary learning} \cite{olshausen1996,OLSHAUSEN1997,DL_Kreutz03}, one wants to find the two factors $\mathbf{A}$, of size $N\times P$, and $\mathbf{B}$, of size  $P\times M $, such that the columns of $\mathbf{A}$ form an overcomplete basis (i.e. $P> N$) and $\mathbf{B}$ is sparse. In other words, one aims for a sparse representation of the data $\mathbf{Y}$, a common strategy adopted for the denoising of large matrices which aims at reconstructing $\mathbf{Y}^*$ from the measured $\mathbf{Y}$ \cite{repr_lerning_review}.
The much studied Boltzmann machines are another instance of this approach \cite{BM_Hinton,Sala_RBM}: given a set of observations on their visible units, after training they create an internal representation on the hidden units. The core idea of representation learning, and presumably the reason of its effectiveness \cite{Mairal_colorrestoration,Sapiro_review,Image_denoising}, is the extraction of characteristic features, \emph{e.g.} the overcomplete basis in $\mathbf{A}$, that if properly recombined can reconstruct the data, or even some of their missing parts, as in recommender systems \cite{PMF_salakhutdinov}. 
Another motivation for studying  matrix factorization is to find new ways of training deep networks. Given a desired output, the process of learning can be decomposed layer by layer, in which the task is to find a set of synaptic weights together with the internal representation of the data in the previous layer. This is a matrix factorization task (complicated by the non-linearity), that one can hope to turn into a self-consistent solution of deep network training, following what was done for multi-layer generalized linear estimation \cite{MKMZ}.

In the present paper we propose a new way to study matrix factorization, in the difficult regime where the dimensions of the factors go to infinity simultaneously, by mapping it to neural network models used for associative memories. While our analysis can be carried out for the generic problem, for the sake of clarity we present it here in its symmetric version and with Gaussian noise. In this case one measures the $N\times N$ matrix
\begin{align}\label{eq:dic_channel}
\mathbf{Y}=\frac{\boldsymbol{\xi}\boldsymbol{\xi}^\intercal
    }{\sqrt{N}}+\sqrt{\Delta}\mathbf{Z}
\end{align}
built from the factors $\boldsymbol{\xi}=(\xi_i^\mu)$, $1\leq i\leq N$, $1\leq \mu \leq P$.  Here $\mathbf{Z}$ is a symmetric noise matrix in which the $i\leq j$ elements are all independent and Gaussian, with mean $0$ and variance $1+\delta_{ij}$. We shall refer to the vectors $\boldsymbol{\xi}^\mu=(\xi_i^\mu)$, $1\leq i\leq N$, as \emph{patterns}.

When the rank $P$ is finite, the model \eqref{eq:dic_channel} goes under the name of spiked Wigner model, introduced in \cite{Johnstone_WSM} as a model for Principal Component Analysis (PCA), and studied in detail in recent years \cite{lesieur2015mmse,channel_universality,Barbier_2019,ElAlaoui,BBP,BENAYCHGEORGES2011}

A common setting to tackle \eqref{eq:dic_channel} is the Bayes-optimal one, in which we suppose to know the noise variance $\Delta$ as well as the probabilistic model of the data, including the distribution of the patterns $P_\xi$.
The inference task amounts to reconstructing the patterns stored in $\boldsymbol{\xi}$ from the posterior probability $(1/\mathcal{Z}) P_{\text{prior}}(\mathbf{x}) P(\mathbf{Y}|\mathbf{x})$. 
In presence of Gaussian noise the latter can be seen as a Boltzmann measure at temperature $\Delta$ for a $N\times P$ matrix $\mathbf{x}$ with prior distribution $P_{\text{prior}}(\mathbf{x})$, and energy
\begin{align}\label{energy_Bayes}
E(\mathbf{x}|\mathbf{Y})= \frac{1}{2} \sum_{i,j=1}^N \Big(Y_{ij}-\sum_{\mu=1}^P\frac{x_i^\mu x_j^\mu}{\sqrt{N}}\Big)^2\,.
\end{align}
Bayes optimality means that the prior measure on $\mathbf{x}$, $ P_{\text{prior}}(\mathbf{x})$, is the same measure as $P_\xi$ from which the original patterns were drawn. It is well known that a typical sample from the Bayes-optimal Boltzmann measure gives the smallest possible reconstruction error, in average over the observations $\mathbf{Y}$. This gives an information-theoretical bound for the performance of reconstruction algorithms. For finite $P$, these benchmark values were computed in the aforementioned literature in the thermodynamic limit $N\to\infty$, and there exist polynomial-time algorithms known as Approximate Message Passing (AMP) \cite{KabashimaAMP,donohoAMP} able to saturate the Bayes-optimal performances. 

Much less is known in the regime of extensive rank, namely when $P$ and $N$ go to infinity with fixed $P/N =\alpha$. If one restricts to the simpler question of denoising, a possible strategy, introduced in \cite{RIE_bouchaud}, is that of using \emph{Rotationnally Invariant Estimators} (RIE) for the hidden matrix $\boldsymbol{\xi}\boldsymbol{\xi}^T/\sqrt{N}$. 
RIEs produce an estimation of $\mathbf{Y}^*$ that has the  same eigenbasis as the observed $\mathbf{Y}$. The task then reduces to denoising the eigenvalues of $\mathbf{Y}$. Notice however that RIEs are not Bayes-optimal if the prior $P_\xi$ is not rotational invariant.

Previous attempts at solving matrix factorization in the extensive rank regime \cite{Marc-Kabashima,perturbative_Maillard21,barbier2022DL} focused on the Bayes-optimal Boltzmann measure with energy (\ref{energy_Bayes}). This turns out to be a difficult matrix model with quenched randomness that has defied studies so far. For instance, it could not be established if a solution exists in the extensive rank case.

We propose here an alternative strategy: considering $\boldsymbol{\xi}$ as composed of $P$ patterns $\boldsymbol{\xi}^\mu$, $\mu=1,2,\dots,P$, we aim for one of them at a time. Assuming that we are able to get efficiently an estimate $\boldsymbol{\eta}^P$ of a first pattern, say $\boldsymbol{\xi}^P$, then we can build a rank-one contribution and subtract it from $\mathbf{Y}\equiv\mathbf{Y}_0$, obtaining $\mathbf{Y}_1=\mathbf{Y}_0-\boldsymbol{\eta}^P\boldsymbol{\eta}^{P\intercal}/\sqrt{N}$. Then we iterate this process $P$ times. We will refer to this iterative scheme as \emph{decimation}. 
Notice that decimation bears some similarity to the `unlearning' procedure studied in \cite{Hopfield_unlearning,vanHemmen_unlearning,Benedetti_unlearning}. However, while unlearning aims at smoothing the landscape, and increasing the basins of attraction of patterns, by subtracting $\mathbf{Y}_1=\mathbf{Y}_0-\varepsilon \boldsymbol{\eta}^P\boldsymbol{\eta}^{P\intercal}/\sqrt{N}$ with a small $\varepsilon$, here we use $\epsilon=1$ in order to fully erase the found pattern before the next decimation step.

After $R$ steps of decimation, and choosing indices such that approximate values of the last $R$ patterns have been found, we must factorize the reduced matrix 
$\mathbf{Y}_R=\mathbf{Y}-\sum_{\mu=P-R+1}^P \boldsymbol{\eta}^\mu\boldsymbol{\eta}^{\mu\intercal}/\sqrt{N}$. We then try to find a pattern which is a single $N$-component vector $\mathbf{x}=(x_i)$, $1\leq i \leq N$,  with an energy function that can be obtained from \eqref{energy_Bayes}, replacing $\mathbf{Y}$ with $\mathbf{Y}_R$:
\begin{align}\label{hamiltonian}
    \begin{split}
        -E(\mathbf{x}|\mathbf{Y}_R)=\frac{ \sqrt{\Delta}}{2\sqrt{N}}\sum_{i,j=1}^N Z_{ij}x_ix_j+\frac{1}{2 N}\sum_{\mu=1}^P\left(\sum_{i=1}^N\xi_i^\mu x_i\right)^2-\frac{1}{2 N} \sum_{\mu=P-R+1}^P\left(\sum_{i=1}^N\eta_i^\mu x_i\right)^2
        -\frac{\Vert\mathbf{x}\Vert^4}{4 N}\,.
    \end{split}
\end{align}
We aim at sampling a pattern from the Boltzmann distribution 
\begin{align}
\frac{1}{\mathcal{Z}_R} P_\xi(\mathbf{x}) e^{-\beta E(\mathbf{x}|\mathbf{Y}_R)} 
\label{Boltzmann-measure}
\end{align}
at large $\beta$ in the case of factorized priors $P_\xi(\mathbf{x})=\prod_{i\leq N}P_\xi(x_i)$. Expectations w.r.t. \eqref{Boltzmann-measure} are denoted by $\langle\cdot\rangle_R$.
If we choose $P=1$ in \eqref{hamiltonian}, and we decide for finite temperature sampling at $\beta=1/\Delta$, we recover the Bayes-optimal setting for the rank-1 spiked Wigner model.

\begin{figure*}[ht!!!]
    \centering
     \subfloat[$\beta=1/\Delta$, $\rho=1$]{\includegraphics[width=8.6cm,height=5cm]{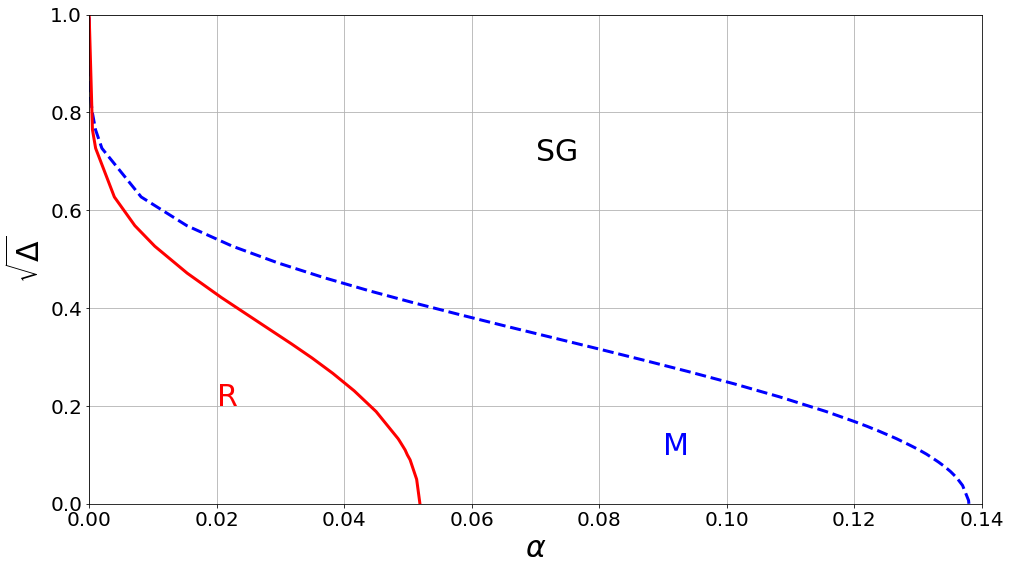}\label{fig:phase_diag_Ising}}\hspace{.1cm}
     \subfloat[$\beta\to\infty$, different values of $\rho$: red and blue $\rho=1$, magenta and cyan $\rho=0.1$, green and yellow $\rho=0.05$.]{\includegraphics[width=8.6cm,height=5cm]{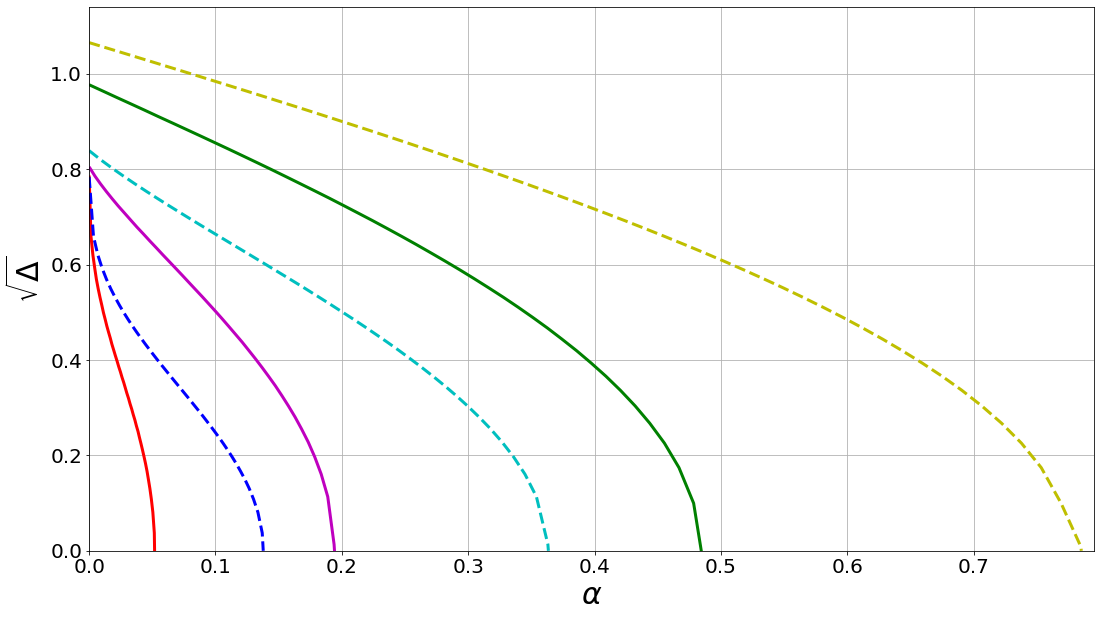}\label{fig:0Tphase_diag}}
     \caption{Phase diagrams for the indicated values of inverse temperature $\beta$ and sparsity $\rho$. (a): the ``R'' phase is the retrieval phase. In the \enquote{M} phase the retrieval states exist, but they are metastable. In the \enquote{SG} region the system is in the spin glass phase, where there is no retrieval. We do not show here the paramagnetic phase which appears above $\Delta=1$. (b): dashed and solid lines refer to the same transitions as in (a). Different colors refer to different sparsities.}
\end{figure*}

The similarity of \eqref{hamiltonian} with the energy of Hopfield's model \cite{Hopfield82} for associative memory helps to get some insight on the physical behaviour of decimation.
In order to explain it, one can start from the ideal case where the last $R$ patterns have been reconstructed exactly ($\boldsymbol{\eta}^\mu=\boldsymbol{\xi}^\mu$ for $\mu>P-R$). Then the energy (\ref{hamiltonian})
favours  the $\mathbf{x}$-configurations that are most aligned to one of the remaining $P-R$ patterns $\boldsymbol{\xi}^\mu$. 
Decimation may work if the minima of the energy \eqref{hamiltonian} are \enquote{retrieval states} close to the $P-R$ patterns that one wants to reconstruct.
In practice, the estimates $(\boldsymbol{\eta}^\mu)_{\mu\geq P-R+1}$ have an $O(N)$ projection onto the related patterns $\boldsymbol{\xi}^\mu$. The $R$ quadratic terms involving the $\boldsymbol{\eta}^\mu$ then repel the $\mathbf{x}$'s from the previously estimated patterns, as wanted. The $\Vert\mathbf{x}\Vert^4$ term simply penalizes large norms.

We stress that in the decimation energy \eqref{hamiltonian} there are three sources of noise, that might disturb pattern retrieval. The first one is  \emph{(a)} the original Gaussian noise $\mathbf{Z}$. Then there is \emph{(b)} a noise due to pattern interferences, the well known limiting factor for patterns' memorization in the Hopfield model \cite{Amit0,Amit1}. 
And thirdly, \emph{(c)} the decimation itself:  the retrieval states $\boldsymbol{\eta}^\mu$ are blurred versions of the $\boldsymbol{\xi}^\mu$. 

Along decimation, the interference noise \emph{(b)} decreases with the number of memorized patterns, but the decimation noise \emph{(c)} increases. 
In order to study the theoretical performance of decimation, we need to identify which of these two mechanisms dominates.

The decimation noise contribution depends on the precision of the retrieval of previous patterns.  Using knowledge from Hopfield's model phenomenology, we shall assume that for large $N$ the retrieved states $(\boldsymbol{\eta}^\mu)_{\mu\geq P-R+1}$ are sampled from a factorized distribution, with an effective local field that is Gaussian centered at $m^\mu \boldsymbol{\xi}^\mu$ with variance $r$, as described in Appendix \ref{app:retrieval}. The crucial parameter $m^\mu$ characterizes the retrieval accuracy, i.e. how close the retrieved state $\boldsymbol{\eta}^\mu$ is to the stored pattern $\boldsymbol{\xi}^\mu$, and it must be determined self-consistently together with the variance $r$.

The retrieval accuracy $m^{P-R}$ on the pattern retrieved using the measure \eqref{Boltzmann-measure} depends on the whole set of previous retrieval accuracies $m^{P-R+q}$, $q=1,...,R$. $m^{P-R}$ can be obtained using the replica method \cite{MPV} which allows to compute the average of $\Phi=(1/N)\log\mathcal{Z}_R $ over the distribution of patterns and of retrieved states, in the large $N$ limit. We use a \emph{Replica Symmetric} (RS) ansatz, which is known to provide a good approximation to the exact result in neural networks \cite{Amit1}. Defining $t=R/P$, we show in Appendix \ref{app:replica-computation} that the free entropy $\Phi$ can be written  in terms of three order parameters: the sought retrieval accuracy $m$, the overlap $q = \mathbb{E}_{\boldsymbol{\xi},\boldsymbol{\eta}}\langle x_i\rangle_R^2$, and the variance $r$. The dependence on $m$ is simple:
\begin{align}\label{RS_free_entropy}
        \begin{split}
        &\Phi=\text{Extr}\Big\{\Phi_0(\alpha,m^{[0,t]};q,r,u)
        -\beta\frac{m^2}{2}+\mathbb{E}_{Z,\xi}\log\int dP_\xi(x)\exp\left(\left(Z\sqrt{r}+\beta m\xi\right)x-\frac{u+r}{2}x^2\right)
        \Big\}
        \end{split}
\end{align}
where $u$ is an auxiliary variational parameter chosen to impose $v=\mathbb{E}_{\boldsymbol{\xi},\boldsymbol{\eta}}\langle \Vert\mathbf{x}\Vert^2\rangle_R/N=1$, and $m^{[0,t]}$ stands for the collection of the accuracies of the previous decimation steps $\{m(\tau),\tau\leq t\}$. 
The value of $m$ where $\Phi$ reaches its maximum is the $R+1$-th retrieval accuracy $m^{P-R}$. It depends crucially on the noise strength $r$ which is the sum of three contributions, $r=r_a+r_b+r_c$ corresponding to the three sources of noise discussed above:
\begin{align}
 r_a &= \beta^2\Delta q\,, \quad r_b= \frac{(1-t)\alpha \beta^2 q}{(1-\beta(1-q))^2}\,,\\
 r_c&=2 \alpha t \beta^2q\int_0^t d\tau \, f(\tau)[1+2\beta^2(1-q)^2 f(\tau)]\,.
\end{align}
$r_c$ is the decimation-induced noise level, it depends on  $f(\tau)= [1-m(\tau)^2]/[1-\beta^2(1-q)^2(1-m(\tau)^2)]$. The other consistency equations are deferred to Appendix \ref{app:FPeqs}, whereas in Appendix \ref{app:0Tlimit} we compute their low temperature limit.

We shall study the performance of decimation assuming the existence of an ``oracle'' algorithm which could be of two types: 1) ``Sampling oracle'': returns a typical configuration sampled from the Boltzmann distribution \eqref{Boltzmann-measure} at inverse temperature $\beta=1/\Delta$; 2) ``Ground state oracle'': returns the lowest energy configuration of energy (\ref{hamiltonian}) (equivalent to sampling at $\beta=\infty$). These oracles will return retrieval states correlated with the stored patterns only in some part of the phase diagram controlled by the parameters $\alpha,\Delta,t$.

\begin{figure}[ht!!!]
    \centering
    \includegraphics[width=8.6cm,height=4.8cm]{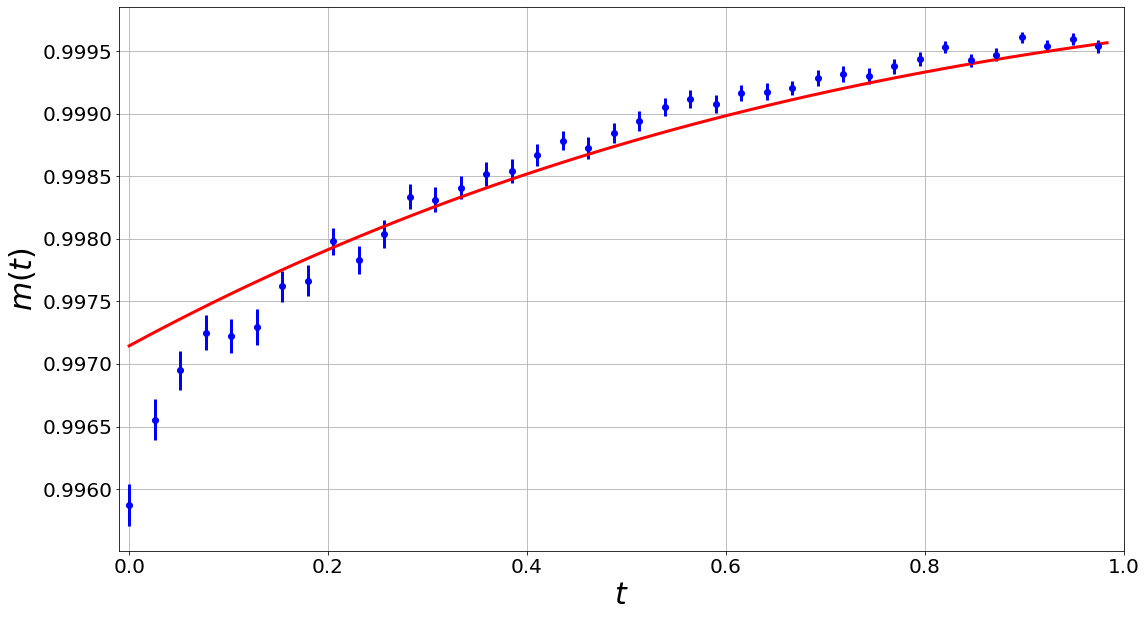}
    \caption{The accuracy of pattern retrieval is plotted vs the decimation parameter $t=R/P$. The red curve is the theoretical prediction with $\alpha=0.03$, $\Delta=0.08$ at $\beta=\infty$. The blue data points are an average over 300 runs of the ground state oracle for $N=1300$, with one standard deviation error-bars.}
    \label{fig:mags_vs_decstep}
  \end{figure}

We start with the first step of decimation, $t=0$, using for definiteness a measure $P_\xi(x)=(1-\rho)\delta_{x,0}+\rho/2 (\delta_{x,1/\sqrt{\rho}}+\delta_{x,-1/\sqrt{\rho}})$, where $\rho$ controls the sparsity. Fig.\ref{fig:phase_diag_Ising} shows the phase diagram for binary patterns with a sampling oracle. The retrieval region R, $\alpha<\alpha_c(\Delta)$, is where the retrieval states dominate the measure. In the M phase, $\alpha_c(\Delta)<\alpha<\alpha_M(\Delta)$, retrieval states are metastable, and in the spin glass SG phase there are no states correlated with the patterns. This is strongly reminiscent of the phase diagram of the Hopfield model \cite{Amit1} and in fact the boundaries of the phase transition lines at $\Delta=0$ are found at the same values, $\alpha_c(0)=0.051$ and $\alpha_M(0)=0.138$. Here $\Delta$ plays a role similar to the temperature in Hopfield model. Fig.\ref{fig:0Tphase_diag} shows that sparsity increases the size of the retrieval phase. It can be actually proved that $\alpha_M(\Delta=0)\to\infty$ as $\rho\to 0$.

As decimation proceeds, the retrieval boundary $\alpha_c$ moves to the left because of the increase of retrieval noise $r_c$, and at the same time the effective $\alpha$ decreases. It turns out that, in all the cases that we have studied, the second effect is more important: if we start the decimation at $t=0$ in the retrieval phase, then the system remains in the retrieval phase for the whole decimation process. This is the crucial point that makes this approach efficient for matrix factorization.
It is illustrated in Fig.\ref{fig:mags_vs_decstep}, where we plot the theoretical value of the retrieval accuracy $m$ at decimation step $R=t P$, computed from \eqref{RS_free_entropy}, versus $t$. It starts from $m=.9971$ at $t=0$ and increases to $m=.9996$ at the end of decimation, showing that the precision of pattern retrieval increases during decimation.

\begin{figure*}[ht!!!]
\centering
     \subfloat[$N=2000$, $P=60$, $\rho=1$.]{\includegraphics[width=8.6cm,height=5cm]{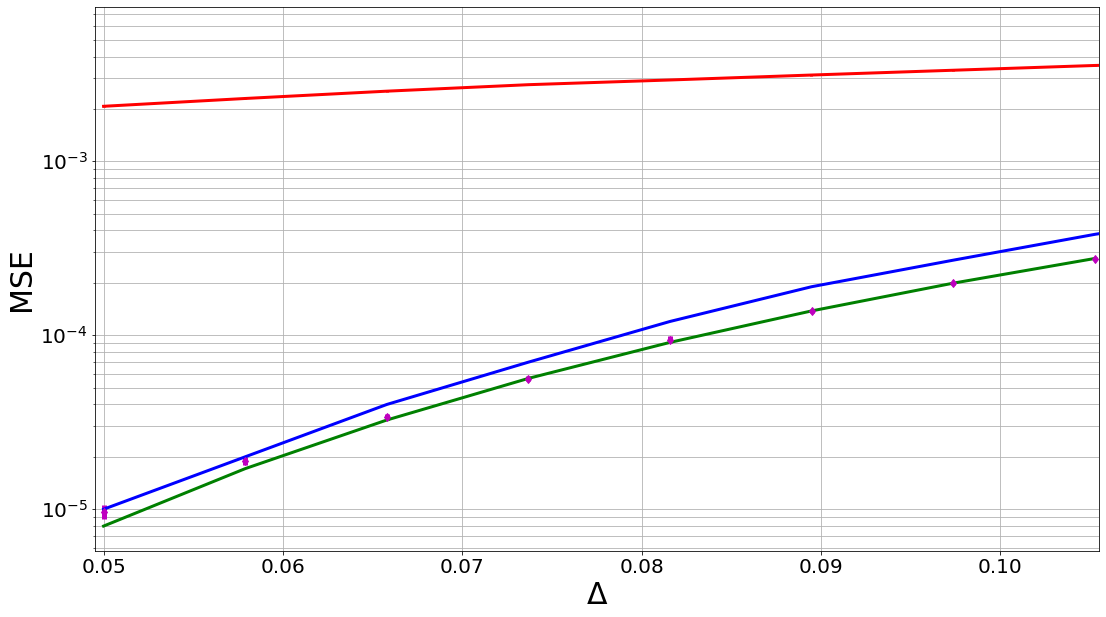}\label{decimation_BABP_Ising}}\hspace{.1cm}
     \subfloat[$N=1000$, $P=400$, $\rho=0.05$.]{\includegraphics[width=8.6cm,height=5cm]{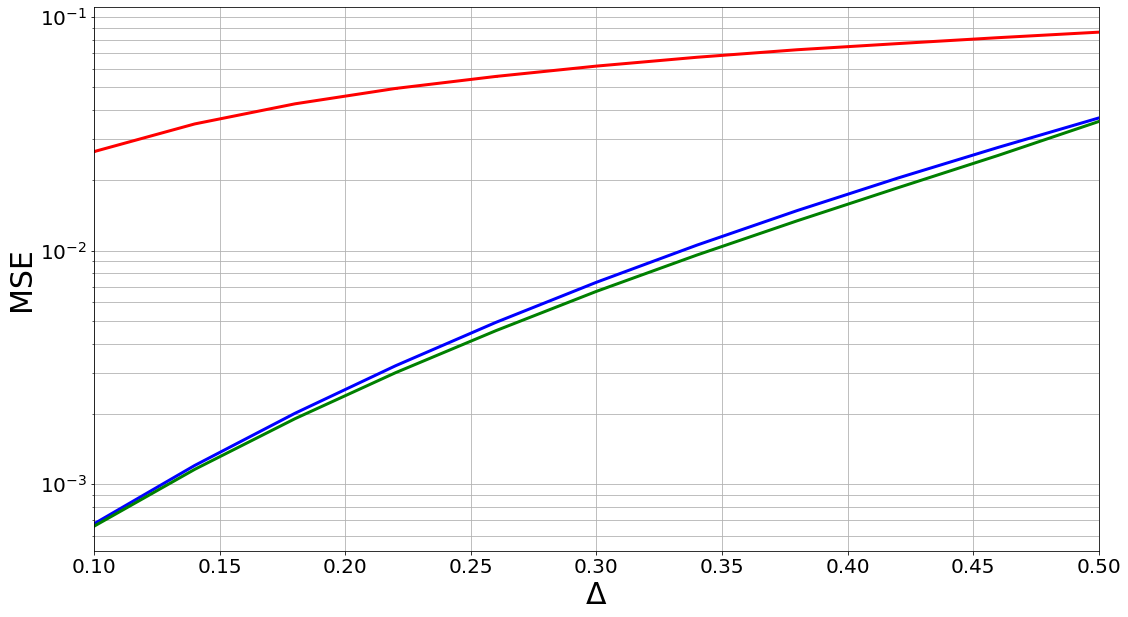}\label{decimation_BABP_sparse}}
      \caption{Matrix denoising: comparison between rotation invariant (red) and denoising based on decimation with a sampling oracle at $\beta=1/\Delta$ (blue) or with a ground state oracle at $\beta=\infty$ (green). Performances are measured by the average mean square error over 30 samples for each $\Delta$. Error bars are too small to be seen.  The magenta points in (a) are obtained via our ground state oracle; error bars are 1 standard deviation of 20 MSEs. The first 5 points are obtained with $N=1800$, the remaining three with $N=1700,1600,1500$ respectively.}
\end{figure*}

We now study the performance of decimation for denoising. The efficiency of a denoiser is measured by the matrix mean square error between the denoiser estimation $\hat{\mathbf{S}}$ and the initial signal $\boldsymbol{\xi}\boldsymbol{\xi}^\intercal/\sqrt{N}$:
\begin{align} \text{mMSE}=\frac{1}{2N^2}\mathbb{E}\Big\Vert\hat{\mathbf{S}}-\frac{\boldsymbol{\xi}\boldsymbol{\xi}^\intercal}{\sqrt{N}}\Big\Vert^2\xrightarrow[]{N\to\infty}\alpha-\alpha\int_0^1d\tau\,m(\tau)^2.
\end{align}
The decimation estimator is $\hat{\mathbf{S}} =\sum_{\mu=1}^P\frac{\boldsymbol{\eta^\mu}\boldsymbol{\eta}^{\mu\intercal}}{\sqrt{N}}$. 
In \figurename\,\ref{decimation_BABP_Ising} and \figurename\,\ref{decimation_BABP_sparse} we show that the decimation denoiser, which exploits the structure of the prior is better than the RIE denoiser.

So far we have established the theoretical performance of decimation based on an oracle able to find retrieval states close to patterns in an associative memory model. However, this retrieval is far from trivial: the oracle cannot use the property of associative memory as we have no knowledge of the patterns.
In a neurobiological perspective, what we seek is an algorithm able to retrieve the stored information in a neural network, knowing the connectivity and synaptic efficacies. In this context, a message passing algorithm for finite rank matrix factorization was tried recently, but it seems to be limited to the case where the rank scales sub-linearly with the number of neurons \cite{GoldtKrzakalaBrunelZdeborova2021}.

We have tested a simple oracle, whose pseudocode can be found in Appendix \ref{app:GSoracle}, based on a Monte Carlo method with simulated annealing starting from infinite temperature (a random configuration) and decreasing it to $T=0$. An extra feature is a `restart' procedure: when the algorithm finds a configuration $x^*$, its energy $E(x^*)$ is compared to the ground state energy $E_{GS}$ at infinite $N$, that we know analytically from the replica method. When $E(x^*)>\theta E_{GS}$, with $\theta$ a threshold initially fixed to $1$, $x^*$ is not accepted as a potential pattern and the procedure is restarted from infinite temperature, with a slightly lower threshold. We have found that the number of restarts needed to find the patterns grows exponentially with $N$. However, although this algorithm needs an exponential time, the pre-factor is small enough that it gives good performance up to $N$ of order $1500$ to $2200$ (depending on the noise level $\Delta$).
It is quite possible that better oracles can be found, without the constraint of satisfying detailed balance.

From our analysis, we are thus able to conclude that \emph{(i)} symmetric matrix factorization is possible in the extensive rank case,  \emph{(ii)} decimation succeeds at exploiting the structure of the signal in the prior and get a better denoiser than RIEs. Furthermore, \emph{(iii)} strong sparsity helps in the reconstruction, making the system more robust against the three noise sources. All these results require to have an oracle able to find a retrieval state, given a matrix of coupling in an associative memory network. We saw that simulated annealing with restarts, although it is an exponential algorithm, provides such an oracle for small enough sizes. Finding a polynomial time oracle would be extremely interesting, both for matrix factorization and neuro-physiological analysis.  Finally, we stress that decimation can be analysed similarly for the asymmetric matrix factorization problem of $\mathbf{Y}$ into $\mathbf{A}\mathbf{B}$ and it will be the object of future work.

\vspace{4pt}

\begin{acknowledgments}
We thank Florent Krzakala, Lenka Zdeborov\'a and Jean Barbier for many fruitful discussions. FC acknowledges financial support from University of Bologna.
\end{acknowledgments}

\newpage

\appendix

\section{Retrieval states and retrieval error}\label{app:retrieval}
\noindent Let us define the Mattis magnetizations
\begin{align}
    &m^\mu(\mathbf{x})=\frac{1}{N}\sum_{i=1}^N\xi_i^\mu x_i\,,\quad\mu=1,\dots,P\\
    &p^\mu(\mathbf{x})=\frac{1}{N}\sum_{i=1}^N\eta_i^\mu x_i\,,\quad\mu=P(1-t)+1,\dots, P\,.
\end{align}
The energy then takes the useful form
\begin{align}\label{Hamiltonian_useful}
    -E(\mathbf{x}|\mathbf{Y_R})=\frac{ \sqrt{
    \Delta}}{2\sqrt{N}}\sum_{i,j=1}^N Z_{ij}x_ix_j+\frac{  N}{2}\sum_{\mu=1}^P(m^\mu(\mathbf{x}))^2-\frac{  N}{2}\sum_{\mu=P(1-t)+1}^P(p^\mu(\mathbf{x}))^2
    -\frac{ \Vert\mathbf{x}\Vert^4}{4 N}+\frac{ \hat{v}}{2}(N-\Vert\mathbf{x}\Vert^2)\,.
\end{align}
The Lagrange multiplier $\hat{v}$ has been introduced to fix $\mathbb{E}_{\boldsymbol{\xi},\boldsymbol{\eta}}\langle \Vert\mathbf{x}\Vert^2\rangle_R/N\xrightarrow[]{N\to\infty}1$. We assume that the oracle will produce $\eta^\mu$ with an asymptotic measure given by
\begin{align}
    \label{local_measure}
    \eta_i^\mu\,\sim\,\langle\cdot\rangle_{\xi_i^\mu,Z}=\frac{\int dP_\xi(x)e^{(Z\sqrt{r}+\beta m^\mu\xi_i^\mu)x
    -\frac{r+u}{2}x^2}(\cdot)}{\int dP_\xi(x)e^{(Z\sqrt{r}+\beta m^\mu\xi_i^\mu)x
    -\frac{r+u}{2}x^2}}\,,\quad \xi_i^\mu\sim P_\xi\,, Z\sim\mathcal{N}(0,1)\text{ independent of other noises}\,,
\end{align}where $m^\mu$, \emph{i.e.} the retrieval accuracy for $\boldsymbol{\eta}^\mu$, and $\,r,\,u$ must be determined self-consistently. Define for later convenience the quantities
\begin{align}
    &\mathbb{E}_{\boldsymbol{\eta}|\boldsymbol{\xi}} [\eta_i^\mu]=m_i^\mu\,,\quad \mathbb{E}_{\boldsymbol{\eta}|\boldsymbol{\xi}}[(\eta_i^\mu)^2]=v_i^\mu\,.
\end{align}Then \eqref{local_measure} has the following implications:
\begin{align}\label{oracle_properties_general}
    &\mathbb{E}_{\boldsymbol{\xi}}[\eta_i^\mu]=\mathbb{E}_{\boldsymbol{\xi}}\mathbb{E}_{\boldsymbol{\eta}|\boldsymbol{\xi}} [\eta_i^\mu]=0\,,\quad \mathbb{E}_{\boldsymbol{\xi}}[\xi_i^\mu m_i^\nu]=m^\mu\delta_{\mu,\nu}\,,\quad \mathbb{E}_{\boldsymbol{\xi}}[v_i^\mu]=1
\end{align}
that are self-consistent with the fixed point equations for each decimation step.

\section{Replica symmetric free entropy}\label{app:replica-computation}
In this section we compute the large $N$ limit of the free entropy
\begin{align}\label{free_entropy_def}
    \Phi_N=\frac{1}{N}\mathbb{E}\log\int dP_\xi(\mathbf{x})\exp\left[-\beta E(\mathbf{x}|\mathbf{Y_R})\right]\,,
\end{align}
where $\mathbb{E}$ is taken w.r.t. all the disorder: $\mathbf{Z},\boldsymbol{\xi},\boldsymbol{\eta}$. This is done using the \emph{replica method} \cite{MPV}.
We thus introduce
\begin{align}
    \label{replicated_annealed_partfunc}
    \mathbb{E}\mathcal{Z}^n_N:=\mathbb{E}_{\mathbf{Z}}\mathbb{E}_{\boldsymbol{\xi},\boldsymbol{\eta}}\int\prod_{a=1}^n dP_\xi(\mathbf{x}_a)\exp\left[-\beta\sum_{a=1}^nE(\mathbf{x}_a|\mathbf{Y_R})\right]\,.
\end{align}
We decompose this computation and start with the first noise terms in \eqref{Hamiltonian_useful}, and the related $\mathbb{E}_\mathbf{Z}$ average
\begin{multline}
        \mathbb{E}_\mathbf{Z}\exp\left(\frac{\beta\sqrt{\Delta}}{2\sqrt{N}}\sum_{i,j=1}^N Z_{ij}\sum_{a=1}^n x_{a,i}x_{a,j}\right)=
        \exp\left(\frac{\beta^2\Delta}{4N}\sum_{i,j=1}^N \sum_{a,b=1}^n x_{a,i}x_{a,j}x_{b,i}x_{b,j}\right)=\\
        =\exp\left(\frac{N\beta^2\Delta}{4}\sum_{a\neq b}^nQ^2(\mathbf{x}_a,\mathbf{x}_b)+\beta^2\Delta\frac{\Vert\mathbf{x}_a\Vert^4}{4 N}\right)\,.
\end{multline}
Now we take care of the penalizing $p$-terms in \eqref{Hamiltonian_useful}. After replicating, their contribution to the partition function is
\begin{align}
    \begin{split}
        A:=\prod_{\mu=P(1-t)+1}^P\prod_{a=1}^n e^{-\frac{N\beta}{2}(p^\mu(\mathbf{x}_a))^2}=\prod_{\mu=P(1-t)+1}^P\prod_{a=1}^n\int\frac{ds_a^\mu}{\sqrt{2\pi}}e^{-\frac{(s_a^\mu)^2}{2}+i\sqrt{\frac{\beta}{N}}s_a^\mu\sum_{j=1}^N\eta_j^\mu x_{a,j}}\,.
    \end{split}
\end{align}
Notice that, thanks to the introduction of the auxiliary Gaussian variables $(s_{a}^\mu)_{a\leq n,P(1-t)< \mu\leq P}$, the exponential is now decoupled over the particle indices $j$. Consider then the expectation of $A$ w.r.t. $\boldsymbol{\eta}$, given $\boldsymbol{\xi}$ with the assumptions \eqref{oracle_properties_general}:
\begin{align}
    \begin{split}
        \mathbb{E}_{\boldsymbol{\eta}|\boldsymbol{\xi}}[A]&=\prod_{\mu=P(1-t)+1}^P\prod_{a=1}^n\int\frac{ds_a^\mu}{\sqrt{2\pi}}e^{-\frac{(s_a^\mu)^2}{2}}\prod_{i=1}^N\mathbb{E}_{\eta^\mu_i|\xi^\mu_i}\exp\left(i\sqrt{\frac{\beta}{N}}\eta_i^\mu\sum_{a=1}^n s_a^\mu x_{a,i}\right)\\
        &=\prod_{\mu=P(1-t)+1}^P\prod_{a=1}^n\int\frac{ds_a^\mu}{\sqrt{2\pi}}\exp\left(-\frac{(s_a^\mu)^2}{2}+\sum_{i=1}^N\log\mathbb{E}_{\eta^\mu_i|\xi^\mu_i}e^{i\sqrt{\frac{\beta}{N}}\eta_i^\mu\sum_{a=1}^n s_a^\mu x_{a,i}}\right)\,.
    \end{split}
\end{align}
Now we can expand the exponential inside the $\log$ up to second order, the remaining terms will be of sub-leading order and thus neglected in the following:
\begin{multline}
    \mathbb{E}_{\boldsymbol{\eta}|\boldsymbol{\xi}}[A]=\prod_{\mu=P(1-t)+1}^P\prod_{a=1}^n\int\frac{ds_a^\mu}{\sqrt{2\pi}}\exp\left(-\frac{(s_a^\mu)^2}{2}+\sum_{a=1}^nis_a^\mu\sqrt{\frac{\beta}{N}}\sum_{i=1}^Nm_i^\mu x_{a,i}-\frac{\beta}{2}\sum_{a,b=1}^ns_a^\mu s_b^\mu\sum_{i=1}^N\frac{(v_i^\mu-(m_i^\mu)^2)}{N}x_{a,i}x_{b,i}\right)\\
    =\prod_{\mu=P(1-t)+1}^P\prod_{a=1}^n\int\frac{ds_a^\mu}{\sqrt{2\pi}}\exp\left[-\frac{1}{2}\sum_{a,b=1}^ns_a^\mu s_b^\mu\left(\delta_{ab}+\beta\sum_{i=1}^N\frac{(v_i^\mu-(m_i^\mu)^2)}{N}x_{a,i}x_{b,i}\right)+\sum_{a=1}^n is_a^\mu\sqrt{\frac{\beta}{N}}\sum_{i=1}^Nm_i^\mu x_{a,i}\right]\,.
\end{multline}
To continue, we  assume condensation on a finite number of patterns, say the first $k$. We focus now on the remaining ones, namely for $\mu>k$:
\begin{align}
    B:=\exp\left[\frac{\beta N}{2}\sum_{a=1}^n\sum_{\mu=k+1}^P(m^\mu(\mathbf{x}_a))^2\right]
    =\int\prod_{\mu=k+1}^P\prod_{a=1}^n\frac{dz_a^\mu}{\sqrt{2\pi}}\exp\left[-\sum_{a=1}^n\sum_{\mu=k+1}^P\frac{(z_a^\mu)^2}{2}+\sqrt{\frac{\beta}{N}}\sum_{a=1}^n\sum_{\mu=k+1}^P z_a^\mu\sum_{i=1}^N x_{a,i}\xi_i^\mu\right]\,.
\end{align}
Putting $A$ and $B$ together, their overall average over $(\boldsymbol{\xi}^{\mu})_{\mu>k}$ takes the form
\begin{multline}
    \mathbb{E}_{(\boldsymbol{\xi}^{\mu})_{\mu>k}}[AB]=\int\prod_{\mu=P(1-t)+1}^P\prod_{a=1}^n\frac{ds_a^\mu}{\sqrt{2\pi}}\int\prod_{\mu=k+1}^P\prod_{a=1}^n\frac{dz_a^\mu}{\sqrt{2\pi}}e^{-\frac{1}{2}\sum_{a=1}^n\left(\sum_{\mu=P(1-t)+1}^P\frac{(s_a^\mu)^2}{2}+\sum_{\mu=k+1}^P\frac{(z_a^\mu)^2}{2}\right)}\\
    \exp\left[
    \sum_{i=1}^N\sum_{\mu=k+1}^P\log\mathbb{E}_{\xi_i^\mu}e^{\sqrt{\frac{\beta}{N}} \sum_{a=1}^nx_{a,i}(\xi_i^\mu z_a^\mu+i\chi(\mu>P-R)m_i^\mu s_a^\mu)-\chi(\mu>P-R)\sum_{a,b=1}^n s_a^\mu s_b^\mu\sum_{i=1}^N \frac{\beta(v_i^\mu-(m_i^\mu)^2) x_{a,i} x_{b,i}}{2 N}}
    \right].
\end{multline}
If we call $\mathbb{E}_{\boldsymbol{\xi}}m_{i}^{\mu\,2}=:\bar{M}^{\mu\,2}$, a further expansion of the exponential yields:
\begin{align}
    \begin{split}
        &\mathbb{E}_{(\boldsymbol{\xi}^{\mu})_{\mu>k}}[AB]=\int\prod_{\mu=P(1-t)+1}^P\prod_{a=1}^n\frac{ds_a^\rho}{\sqrt{2\pi}}\exp\left[-\frac{1}{2}\sum_{\mu=P(1-t)+1}^P\mathbf{s}^\mu\cdot\left(\mathbbm{1}+\beta(1-\bar{M}^{\mu\,2})Q\right)\mathbf{s}^\mu\right]\\
        &\int\prod_{\mu=k+1}^P\prod_{a=1}^{n}\frac{dz_a^\mu}{\sqrt{2\pi}}\exp\left\{-\sum_{\mu=k+1}^P\sum_{a=1}^n\frac{(z_a^\mu)^2}{2}+
        \frac{\beta}{2}\sum_{\mu=k+1}^P\sum_{a,b=1}^{n}
        z_a^\mu z_b^\mu Q(\mathbf{x}_a,\mathbf{x}_b)+\right.\\
        &\left.+i\beta\sum_{\mu=P(1-t)+1}^P\mathbb{E}_{\boldsymbol{\xi}}[\xi_1^\mu m_1^\mu]\sum_{a,b=1}^n z_a^\mu s_b^\mu Q(\mathbf{x}_a,\mathbf{x}_b)-\frac{\beta}{\Delta}\sum_{\mu=P(1-t)+1}^P\sum_{a,b=1}^n(\bar{M}^\mu)^2 s_a^\mu s_b^\mu Q(\mathbf{x}_a,\mathbf{x}_b)
        \right\}
    \end{split}
\end{align}
We can now perform a Gaussian integration over the variables $\mathbf{z}^\mu=(z_a^\mu)_{a\leq n}$:
\begin{align}
    \begin{split}
        \mathbb{E}_{(\boldsymbol{\xi}^{\mu})_{\mu>k}}[AB]&=\int\prod_{\mu=P(1-t)+1}^P\prod_{a=1}^n\frac{ds_a^\rho}{\sqrt{2\pi}}\exp\left[-\frac{1}{2}\sum_{\mu=P(1-t)+1}^P\mathbf{s}^\mu\cdot\left(\mathbbm{1}+\beta Q+\beta^2 Q
        \frac{\mathbb{E}^2_{\boldsymbol{\xi}}[\xi_1^\mu m_1^\mu]}{\mathbbm{1}-\beta Q}
        Q\right)\mathbf{s}^\mu\right]\\
        &\times\exp\left[-\frac{\alpha N}{2}\log\det\left(\mathbbm{1}-\beta Q\right)\right]\,.
    \end{split}
\end{align}
Finally, after an integration over the remaining Gaussian variables $\mathbf{s}^\mu$, and using \eqref{oracle_properties_general}, we get
\begin{align}
    \mathbb{E}_{(\boldsymbol{\xi}^{\mu})_{\mu>k}}[AB]=\exp\left[-\frac{
    \alpha(1-t)N}{2}\log\det\left(\mathbbm{1}-\beta Q\right)-\frac{1}{2}\sum_{\mu=P(1-t)+1}^P\log\det\left(\mathbbm{1}-(1-m^2(\tau^\mu))\beta^2 Q^2\right)\right]\,,
\end{align}
where $\tau^\mu=(1-(\mu-1)/P)$, and $m(\tau^\mu)=m^\mu$ are the previous retrieval accuracies. It remains to analyze the contribution given by $(\boldsymbol{\xi}^\mu)_{\mu\leq k}$:
\begin{align}
    C:=\exp\left[\frac{\beta N}{2}\sum_{a=1}^n\sum_{\mu=1}^k(m^\mu(\mathbf{x}_a))^2\right]=\int\prod_{a=1}^n\prod_{\mu=1}^k dm^\mu_a\sqrt{\frac{\beta N}{2\pi}}\exp\left[\sum_{a=1}^n\sum_{\mu=1}^k\left(-N\beta\frac{(m^\mu_a)^2}{2}+\beta m^\mu_a\sum_{i=1}^N\xi_i^\mu x_{a,i}\right)\right]\,.
\end{align}

Before plugging the contributions coming from $A$, $B$ and $C$ into $\mathbb{E}\mathcal{Z}_N^n$ we need to introduce a collection of Dirac deltas to fix the desired order parameters, that are organized in the overlap matrix $(Q(\mathbf{x}_a,\mathbf{x}_b))_{a,b=1}^n$:
\begin{align}
    1=\int\prod_{a\leq  b\leq n}dq_{ab}\delta(Q(\mathbf{x}_a,\mathbf{x}_b)-q_{ab})=\int\prod_{a\leq b\leq n}\frac{Ndr_{ab}dq_{ab}}{4\pi i}\exp\left[-\frac{1}{2}\sum_{a,b=1}^nr_{ab}(Nq_{ab}-\sum_ix_{a,i}x_{b,i})\right]\,.
\end{align}
Hence, the averaged replicated partition function, at leading exponential order in $N$, takes the form
\begin{align}
    \begin{split}
        \mathbb{E}\mathcal{Z}_N^n&=\int\prod_{a\leq b\leq n}\frac{Ndr_{ab}dq_{ab}}{4\pi i}\int\prod_{a=1}^n\prod_{\mu=1}^k dm^\mu_a\sqrt{\frac{N \beta}{2\pi}}\exp\left[-\frac{N}{2}\sum_{a,b}r_{ab}q_{ab}-\frac{\beta N}{2}\sum_{a=1}^n\sum_{\mu=1}^k(m^\mu_a)^2 \right]\\
        &\times\exp\left[ -\frac{1}{2}\sum_{\mu=P(1-t)+1}^P\log\det\left(\mathbbm{1}-(1-m^2(\tau^\mu))\beta^2Q^2\right)\right]\\
        &\times\exp\left[-\frac{
        \alpha(1-t)N}{2}\log\det\left(\mathbbm{1}-\beta Q\right)+N\beta^2\Delta\sum_{a\neq b,1}^n\frac{q_{ab}^2}{4}+N\beta\sum_{a=1}^n\Big(\frac{\beta\hat{v}}{2}(1-q_{aa})+\frac{\beta\Delta-1}{4}q_{aa}^2\Big)\right]\\
        &\times\left(\int \prod_{\mu=1}^kdP_\xi(\xi^\mu)\prod_{a=1}^ndP_\xi(x_{a})\exp\left[\frac{1}{2}\sum_{a,b=1}^nr_{ab}x_{a}x_{b}+\beta\sum_{\mu=1}^k\sum_{a=1}^n{m^\mu_a}\xi^\mu x_{a}\right]\right)^N\, ,
    \end{split}
\end{align}
where we denote $Q=(q_{ab})_{a,b=1}^n$. We can finally express the replicated free entropy with a variational principle coming from a saddle point argument applied to the formula above:
\begin{align}\label{replicated_free_entropy}
    \begin{split}
        &\Phi_n:=\lim_{N\to\infty}\Phi_{N,n}=\frac{1}{n}\text{Extr}\Big\{-\frac{1}{2}\sum_{a,b}r_{ab}q_{ab}
        -\frac{\beta}{2}\sum_{a=1}^n\sum_{\mu=1}^k(m^\mu_a)^2-\frac{
        \alpha(1-t)N}{2}\log\det\left(\mathbbm{1}-\beta Q\right)\\
        &+\beta\sum_{a=1}^n\Big(\frac{\hat{v}(1-q_{aa})}{2}+\frac{\beta\Delta-1}{4}q_{aa}^2\Big)-\frac{\alpha t}{2R}\sum_{\mu=P(1-t)+1}^P\log\det\left[\mathbbm{1}-(1-m^2(\tau^\mu))\beta^2Q\right]\\
        &+\beta^2\Delta\sum_{a\neq b,1}^n\frac{q_{ab}^2}{4}+\log\int\prod_{\mu=1}^k\mathbb{E}_{\xi^\mu}\int\prod_{a=1}^n dP_\xi(x_a)\exp\left[\frac{1}{2}\sum_{a,b=1}^nr_{ab}x_{a}x_{b}+\beta\sum_{\mu=1}^k\sum_{a=1}^n m^\mu_a \xi^\mu x_{a}\right]
        \Big\}\,.
    \end{split}
\end{align}
The normalized sum over $\mu=P(1-t)+1,\dots,P$ on the second line can be turned into an integral $\int_0^t\,d\tau\dots$ in the large $N$ limit.
The extremization is taken w.r.t. the collection of parameters $(r_{ab},q_{ab})_{a,b=1}^n$, $(m_a^\mu)_{a=1,\mu=1}^{n,k}$. Within the replica symmetric ansatz
\begin{align}
    \begin{cases}
    r_{ab}=r\,,\quad a\neq b\\
    r_{aa}=-u
    \end{cases}\quad 
    \begin{cases}
    q_{ab}=q\,,\quad a\neq b\\
    q_{aa}=v
    \end{cases}\quad
    m_a^\mu=m^\mu\,,\quad 
    Q=\begin{pmatrix}
    v     &q&q&\dots&q\\
    q     &v&q&\dots&q\\
    q     &q&v&\dots&q\\
    \vdots&\vdots&\vdots&\ddots&\vdots\\
    q&q&q&\dots&v
    \end{pmatrix}\in \mathbb{R}^{n\times n}\,.
\end{align}
The determinants of $\mathbbm{1}-\beta Q$ and $\mathbbm{1}-\beta^2(1-m^2(\tau)) Q$ are then easily computed:
\begin{align}
    &\det\left(\mathbbm{1}-\beta Q\right)=\left(1-\beta(v-q)\right)^n\left[1-n\frac{\beta q}{1-\beta(v-q)}\right]\\
    &\det\left(\mathbbm{1}-(1-m^2(\tau))\beta^2Q^2\right)=\left[1-(1-m^2(\tau))\beta^2(v-q)^2\right]^{n-1}\times\left[1-(1-m^2(\tau))\beta^2\left(v+(n-1)q\right)^2\right]\,.
\end{align}
Further simplifications occur for the other terms in the replicated free entropy. In particular the remaining $\log$ integral has to be treated as follows:
\begin{multline}
    \int\prod_{\mu=1}^k\mathbb{E}_{\xi^\mu}\int\prod_{a=1}^n dP_\xi(x_a)\exp\left[\frac{r}{2}\sum_{a\neq b,1}^nx_{a}x_{b}-\frac{u}{2}\sum_{a=1}^nx_a^2+\beta\sum_{\mu=1}^k m^\mu\xi^\mu \sum_{a=1}^n x_{a}\right]=\\
    =\int\prod_{\mu=1}^k\mathbb{E}_{\xi^\mu}\int\prod_{a=1}^n dP_\xi(x_a)\exp\left[\frac{r}{2}\left(\sum_{a=1}^nx_{a}\right)^2-\frac{u+r}{2}\sum_{a=1}^nx_a^2+\beta\sum_{\mu=1}^km^\mu\xi^\mu \sum_{a=1}^n x_{a}\right]=\\
    =\mathbb{E}_Z\int\prod_{\mu=1}^k\mathbb{E}_{\xi^\mu}\prod_{a=1}^n\int dP_\xi(x_a)\exp\left[\sqrt{r}Zx_{a}-\frac{u+r}{2}x_a^2+
    \beta\sum_{\mu=1}^km^\mu\xi^\mu x_{a}\right]=\\
    =\mathbb{E}_Z\mathbb{E}_{\boldsymbol{\xi}}\left[\int dP_\xi(x)\exp\left(\left(Z\sqrt{r}+\beta \mathbf{m}\cdot\boldsymbol{\xi}\right)x-\frac{u+r}{2}x^2\right)\right]^n
\end{multline}where $Z\sim\mathcal{N}(0,1)$, $\boldsymbol{\xi}=(\xi^1,\dots,\xi^k)$, $\mathbf{m}=(m^1,\dots,m^k)$\,. Finally, expanding at first order in $n$ one has:
\begin{align}\label{RS_free_entropy_suppmat}
    \begin{split}
        &\Phi_n:=\text{Extr}\Big\{\frac{rq+uv}{2}
        -\beta\sum_{\mu=1}^k\frac{(m^\mu)^2}{2}-\frac{\beta^2\Delta q^2}{4}-\frac{
        \alpha(1-t)}{2}\left[\log\left(1-\beta(v-q)\right)-\frac{\beta q}{1-\beta(v-q)}\right]\\
        &-\frac{\alpha t}{2}\int_0^td\tau\left[\log\left(1-(1-m^2(\tau))\beta^2\left(v-q\right)^2\right)-\frac{2\beta^2q(v-q)(1-m^2(\tau))}{1-\beta^2(1-m^2(\tau))(v-q)^2}\right]\\
        &\quad+\beta\Big(\frac{\hat{v}(1-v)}{2}+\frac{\beta\Delta-1}{4}v^2\Big)+\mathbb{E}_{Z,\boldsymbol{\xi}}\log\int dP_\xi(x)\exp\left(\left(Z\sqrt{r}+\beta\mathbf{m}\cdot\boldsymbol{\xi}\right)x-\frac{u+r}{2}x^2\right)
        \Big\}+O(n)\,.
    \end{split}
\end{align}
It suffices then to let $n\to0$.

\subsection{Fixed point equations}\label{app:FPeqs}
The stationarity conditions coming from \eqref{RS_free_entropy_suppmat} are
\begin{align}
    &v=1\\
    \label{hatv_equation}
    &v=\mathbb{E}_{\boldsymbol{\xi},\boldsymbol{\eta}}\langle X^2\rangle_{R}\\
    \label{m_equation}
    &m^\mu=\mathbb{E}_{\boldsymbol{\xi},\boldsymbol{\eta}}\xi^\mu \langle X\rangle_{R}\,,\quad \mu=1,\dots, k\\
    \label{q_equation}
    &q=\mathbb{E}_{\boldsymbol{\xi},\boldsymbol{\eta}}\langle X\rangle_{R}^2\\
    \begin{split}\label{r_equation}
    &r=\frac{\alpha(1-t)\beta^2 q}{(1-\beta(v-q))^2}+\beta^2\Delta q+\alpha t\int_0^t\,d\tau\frac{2q\beta^2(1-m^2(\tau))}{1-\beta^2(1-m^2(\tau))(v-q)^2}\left[1+\frac{2\beta^2(v-q)^2(1-m^2(\tau))}{1-\beta^2(1-m^2(\tau))(v-q)^2}\right]
    \end{split}\\
    \begin{split}\label{u_equation}
        &u=\beta\hat{v}+\beta(1-\beta\Delta)v-\alpha(1-t)\beta\frac{1-\beta(v-2q)}{(1-\beta^2(v-q))^2}\\
        &\qquad\qquad-\alpha t\int_0^t\,d\tau\left[\frac{2v\beta^2(1-m^2(\tau))}{1-\beta^2(1-m^2(\tau))(v-q)^2}+q\frac{4\beta^4(v-q)^2(1-m^2(\tau))^2}{(1-\beta^2(1-m^2(\tau))(v-q)^2)^2}\right]\,.
    \end{split}
\end{align}
The first equation, corresponding to $\hat{v}$, can be directly eliminated. Notice that the effect of decimation is visible only in the variables $u$ and $r$ that affect the local measure \eqref{local_measure}. For all practical purposes, we will make finite size simulations and use the discretized form present in \eqref{replicated_free_entropy} of the integral accounting for decimation contributions, starting from step $0$, when no pattern has been retrieved yet. Finally, notice that mixed states solutions are possible, with the estimates aligning to more than $1$ pattern, \emph{i.e.} different $m^\mu$'s in \eqref{m_equation} are non-vanishing. This is not desirable in inference, since one wants to estimate one pattern at a time with the best possible performance.

\subsection{The zero temperature limit}\label{app:0Tlimit}
Let us express the $\beta\to\infty$ limit of the free entropy with a prior of the form
\begin{align}
    P_\xi=(1-\rho)\delta_0+\frac{\rho}{2}\left[\delta_{-1/\sqrt{\rho}}+\delta_{1/\sqrt{\rho}}\right]\,,\quad \rho\in(0,1]\,.
\end{align}
For future convenience we introduce the notations
\begin{align}\label{CUrbarr_definition}
    &C:=\beta(1-q)\,\in[0,1)\,,\quad \bar{r}:=r/\beta^2\,,\quad U:=\frac{u+r}{\beta}
\end{align}where $q$ is intended as the stationary value of the overlap solving the fixed point equations. Denote $\mathbf{m}=(m^\mu)_{\mu=1}^k$, where $k$ is the maximum number of condensed patterns. In the low temperature limit the free entropy, re-scaled by $\beta$, and evaluated at the stationary values of the parameters involved has the form
\begin{align}
    \begin{split}
        \frac{1}{\beta}\Phi&=-\frac{\bar{r}C}{2}+\frac{U}{2}+\frac{\alpha(1-t)}{2(1-C)}-\frac{1}{4}-\frac{\mathbf{m}^2}{2}+\frac{\Delta C}{2}+\psi+\alpha t\int_0^t d\tau \frac{C(1-m^2(\tau))}{1-(1-m^2(\tau))C^2}
    \end{split}
\end{align}
where
\begin{align}\label{psi_definition}
    \psi=\frac{1}{\beta} \mathbb{E}_{\boldsymbol{\xi},Z}\log\left[1-\rho+\rho\cosh\frac{\beta}{\sqrt{\rho}}\left(Z\sqrt{\bar{r}}+ \mathbf{m}\cdot\boldsymbol{\xi}\right)\exp\left(
   -\frac{\beta U}{2\rho}\right)\right]\,.
\end{align}
When $\beta\to\infty$ we have to distinguish two cases in the $Z$ average:
\begin{align}\label{psi_simplified1}
    \begin{split}
        \psi=O\Big(\frac{1}{\beta}\Big)+&\frac{1}{\beta}\mathbb{E}_{\boldsymbol{\xi}}\left(\int_{-\mathbf{m}\cdot\boldsymbol{\xi}/\sqrt{\bar{r}}+U/2\sqrt{\bar{r}\rho}}^\infty+\int^{-\mathbf{m}\cdot\boldsymbol{\xi}/\sqrt{\bar{r}}-U/2\sqrt{\bar{r}\rho}}_{-\infty}\right)\frac{dz\,e^{-\frac{z^2}{2}}}{\sqrt{2\pi}}
        \log\left[1-\rho+\rho\cosh\frac{\beta}{\sqrt{\rho}}\left(z\sqrt{\bar{r}}+ \mathbf{m}\cdot\boldsymbol{\xi}\right)e^{-\frac{\beta U}{2\rho}}\right].
    \end{split}
\end{align}
The $O(\beta^{-1})$ instead comes from integration on the interval $[-\mathbf{m}\cdot\boldsymbol{\xi}/\sqrt{\bar{r}}-U/2\sqrt{\bar{r}\rho},-\mathbf{m}\cdot\boldsymbol{\xi}/\sqrt{\bar{r}}+U/2\sqrt{\bar{r}\rho}]$ of the same integrand, that can be easily bounded.

Let us now focus on the first integral in \eqref{psi_simplified1}. The hyperbolic cosine and the exponential in $U$ dominate on the other terms in the $\log$. Taking into account the exponential growth in the selected range of $z$-values the first integral can be approximated with:
\begin{multline}
        \mathbb{E}_{\boldsymbol{\xi}}\int_{-\mathbf{m}\cdot\boldsymbol{\xi}/\sqrt{\bar{r}}+U/2\sqrt{\bar{r}\rho}}^\infty\frac{dz}{\sqrt{2\pi}}e^{-\frac{z^2}{2}}\left(\frac{Z\sqrt{\bar{r}}+ \mathbf{m}\cdot\boldsymbol{\xi}}{\sqrt{\rho}}-\frac{U}{2\rho}\right)=
        \sqrt{\frac{\bar{r}}{2\pi\rho}}\mathbb{E}_{\boldsymbol{\xi}}e^{-\frac{1}{2\bar{r}}\left(\frac{U}{2\sqrt{\rho}}-\mathbf{m}\cdot\boldsymbol{\xi}\right)^2}+\\
        +\mathbb{E}_{\boldsymbol{\xi}}\left(\frac{ \mathbf{m}\cdot\boldsymbol{\xi}}{\sqrt{\rho}}-\frac{U}{2\rho}\right)\int_{-\mathbf{m}\cdot\boldsymbol{\xi}/\sqrt{\bar{r}}+U/2\sqrt{\bar{r}\rho}}^\infty\frac{dz}{\sqrt{2\pi}}e^{-\frac{z^2}{2}}\,.
\end{multline}
The second integral in \eqref{psi_simplified1} can be treated similarly. Putting all the terms together one gets
\begin{align}\label{re-scaled_free_entropy1}
    \begin{split}
        \frac{1}{\beta}\Phi&=-\frac{\bar{r}C}{2}+\frac{\Delta C}{2}+\frac{U}{2}+\frac{\alpha(1-t)}{2(1-C)}-\frac{1}{4}-\frac{\mathbf{m}^2}{2}+\sqrt{\frac{\bar{r}}{2\pi\rho}}\mathbb{E}_{\boldsymbol{\xi}}\left[e^{-\frac{1}{2\bar{r}}\left(\frac{U}{2\sqrt{\rho}}-\mathbf{m}\cdot\boldsymbol{\xi}\right)^2}+
        e^{-\frac{1}{2\bar{r}}\left(\frac{U}{2\sqrt{\rho}}+\mathbf{m}\cdot\boldsymbol{\xi}\right)^2}\right]+\\
        &+\mathbb{E}_{\boldsymbol{\xi}}\frac{\mathbf{m}\cdot\boldsymbol{\xi}}{2\sqrt{\rho}} \left[\text{erf}\left(\frac{\mathbf{m}\cdot\boldsymbol{\xi}+\frac{U}{2\sqrt{\rho}}}{\sqrt{2\bar{r}}}\right)
        +\text{erf}\left(\frac{\mathbf{m}\cdot\boldsymbol{\xi}-\frac{U}{2\sqrt{\rho}}}{\sqrt{2\bar{r}}}\right)\right]\\
        &-\frac{U}{4\rho}\mathbb{E}_{\boldsymbol{\xi}} \left[2+\text{erf}\left(\frac{\mathbf{m}\cdot\boldsymbol{\xi}-\frac{U}{2\sqrt{\rho}}}{\sqrt{2\bar{r}}}\right)
        -\text{erf}\left(\frac{\mathbf{m}\cdot\boldsymbol{\xi}+\frac{U}{2\sqrt{\rho}}}{\sqrt{2\bar{r}}}\right)\right]+\alpha t\int_0^t d\tau \frac{C(1-m^2(\tau))}{1-(1-m^2(\tau))C^2}\,.
    \end{split}
\end{align}
Using the fact taht all the parameters are evaluated at their stationary values, the previous formula can be further simplified by looking at the limiting version of the fixed point equations. In particular we have that
\begin{align}\label{C_equation}
    C=\frac{1}{\sqrt{2\pi\rho\bar{r}}}\mathbb{E}_{\boldsymbol{\xi}}\left[\exp\left(-\left(\frac{U/2\sqrt{\rho}-\mathbf{m}\cdot\boldsymbol{\xi}}{\sqrt{2\bar{r}}}\right)^2\right)+\exp\left(-\left(\frac{U/2\sqrt{\rho}+\mathbf{m}\cdot\boldsymbol{\xi}}{\sqrt{2\bar{r}}}\right)^2\right)\right]\,.
\end{align}
The value of $\bar{r}$ can be found directly from \eqref{r_equation} by multiplying it by $\beta^{-2}$:
\begin{align}
    \label{barr_equation}
    \bar{r}=\frac{\alpha(1-t)}{(1-C)^2}+\Delta+\alpha t\int_0^t\,d\tau\frac{2((1-m^2(\tau))}{1-(1-m^2(\tau))C^2}\left[1+\frac{2C^2(1-m^2(\tau))}{1-(1-m^2(\tau))C^2}\right]
    \,.
\end{align}
Furthermore, from \eqref{hatv_equation} with $v=1$ one has
\begin{align}
    \label{U_equation_noiseless}
    \mathbb{E}_{\boldsymbol{\xi}}\left[\text{erf}\left(\frac{U/2\sqrt{\rho}-\mathbf{m}\cdot\boldsymbol{\xi}}{\sqrt{2\bar{r}}}\right)+\text{erf}\left(\frac{U/2\sqrt{\rho}+\mathbf{m}\cdot\boldsymbol{\xi}}{\sqrt{2\bar{r}}}\right)\right]=2(1-\rho)\,.
\end{align}
We stress that the l.h.s. of the previous equation is monotonic in $U$, and thus \eqref{u_equation} has a unique solution for fixed sparsity parameter $\rho$.
Finally, from \eqref{m_equation} and \eqref{psi_definition}
\begin{align}\label{m_equation_noiseless}
    \mathbf{m}=\mathbb{E}\boldsymbol{\xi}\langle X\rangle_{Z,\boldsymbol{\xi}}=\frac{\partial\psi}{\partial\mathbf{m}}=\mathbb{E}_{\boldsymbol{\xi}}\frac{\boldsymbol{\xi}}{2\sqrt{\rho}}\left[\text{erf}\left(\frac{\mathbf{m}\cdot\boldsymbol{\xi}-U/2\sqrt{\rho}}{\sqrt{2\bar{r}}}\right)+\text{erf}\left(\frac{U/2\sqrt{\rho}+\mathbf{m}\cdot\boldsymbol{\xi}}{\sqrt{2\bar{r}}}\right)\right]\,.
\end{align}

If we insert these conditions in \eqref{re-scaled_free_entropy1} we get
\begin{align}
    \label{free_entropy_noiseless}
    \frac{\Phi}{\beta}=\frac{\alpha(1-t)}{2(1-C)^2}+\Delta C-\frac{1}{4}+\frac{\mathbf{m}^2}{2}+2\alpha t\int_0^t d\tau \frac{C(1-m^2(\tau))}{[1-(1-m^2(\tau))C^2]^2}\,.
\end{align}

\section{Ground state oracle for binary spins}\label{app:GSoracle}
Here we provide the pseudo-code for the ground state oracle for binary spins, based on an adaptation of simulated annealing. Algorithm \ref{alg:GSO} is itself run multiple times, typically 5 times or more, and we chose the outcome that required the least number of restarts.

\begin{algorithm}[H]
\caption{Ground state oracle with $P_\xi(x)=\frac{1}{2}(\delta_{x,-1}+\delta_{x,1})$}\label{alg:GSO}
\begin{algorithmic}
\Require $N$, $P$ (or $\alpha$), $\mathbf{Y}$, threshold ($\in\mathbb{R}_+$), maxr ($\in\mathbb{N}$)
\State restarts $\gets 0$ 
\State $\mathbf{S}\gets N\times N$ zeros matrix 
\While{$i \leq P$}
    \State $\mathbf{s}$, trials $\gets \text{SA}(N,\mathbf{Y},\text{threshold},\text{maxr},\text{restarts})$ \Comment{See the code of the SA routine below}
    \State restarts $\gets$ restarts$+$trials
    \If{(trials$>299$ \textbf{or} restarts$>$maxr)}
        \State \textbf{break} and start over
    \EndIf
    \State $\mathbf{Y}\gets\mathbf{Y}-\frac{\mathbf{s}\mathbf{s}^\intercal}{\sqrt{N}}$
    \State $\mathbf{S}_i\gets\mathbf{s}$\Comment{$\mathbf{S}_i=i$-th column vector in $\mathbf{S}$}
    \State threshold$\gets$threshold$\,\cdot \,0.9975$
    \State $i\gets i+1$
\EndWhile
\end{algorithmic}
\end{algorithm}

\begin{figure}[ht]
    \centering
    \includegraphics[width=.5\textwidth]{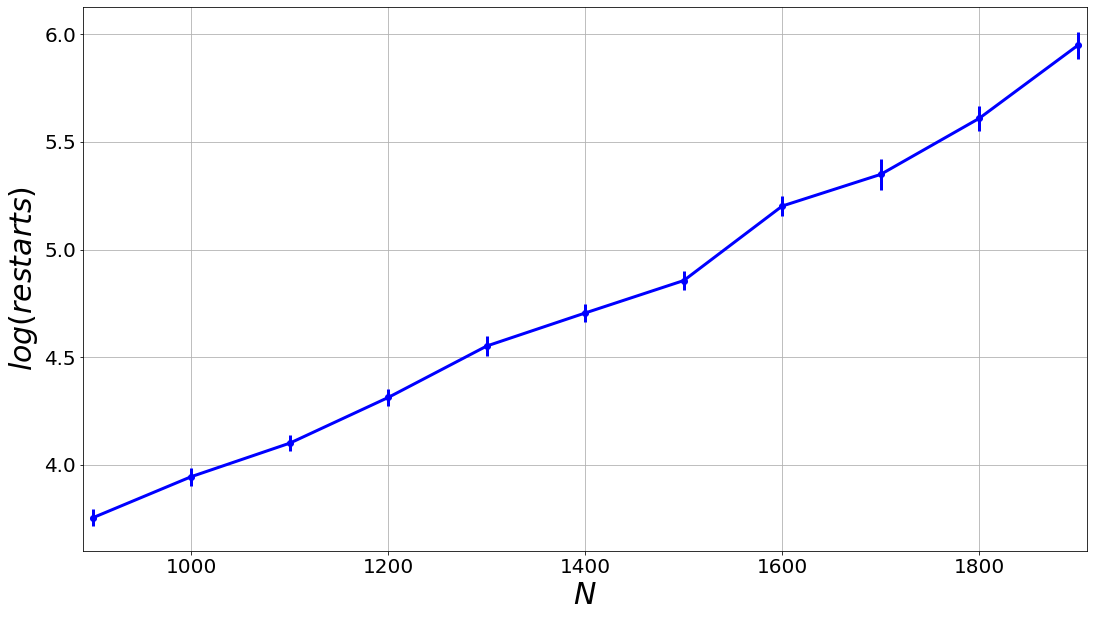}
    \caption{Logarithm of the number of restarts in the whole decimation procedure versus number of signal components for $\Delta=0.05$, $\alpha=0.03$. Recall that $\beta\to\infty$. Each data point is the average of 30 runs, the error bars are one standard deviation.}
    \label{fig:restarts}
\end{figure}

\begin{algorithm}[H]
\caption{Simulated annealing (SA)}\label{alg:SA}
\begin{algorithmic}
\Require $N$, $\mathbf{Y}$, threshold ($\in\mathbb{R}_+$), maxr ($\in\mathbb{N}$), restarts ($\in\mathbb{N}$)
\State itry$\gets 0$
\State found=\textbf{False}
\While{itry$ < 300$}
    \State stop$\gets 0$
    \State $\beta\gets 0$
    \State $\mathbf{s}\gets$ random sample from $\prod_{i=1}^NP_\xi$
    \State itry$\gets$itry$+1$
    \If{itry+restarts$>$maxr}
        \State \Return $\mathbf{s}$, itry
    \EndIf
    \If{itry$\%20=0$}
        \State threshold$\gets$threshold$\,\cdot\,0.9975$
    \EndIf
    
    \While{$k<200$}
        \State $k\gets k+1$
        \State $\beta\gets1+\frac{k}{200}\,\cdot\,99$
        \State $\mathbf{h}\gets \frac{\mathbf{Y}}{\sqrt{N}}\mathbf{s}$
        \State define $ss_i\gets-s_i$ with probability $\frac{1}{1+e^{-2\beta h_i}}$ $\forall \,i$
        \If{$\mathbf{s}-\mathbf{ss}=\mathbf{0}$}
            \State stop$\gets$stop$+1$\Comment{When the search does not move for $5$ consecutive times.}
            \If{stop$>5$}
                \If{$\frac{1}{2}\mathbf{s}^\intercal\frac{\mathbf{Y}}{\sqrt{N}}\mathbf{s}>$threshold}
                    \State \Return $\mathbf{s}$, itry
                \Else 
                    \State\textbf{break}\Comment{wrong energy, try again}
                \EndIf
            \EndIf
        \Else
            \State stop$\gets 0$
            \State $\mathbf{s}\gets\mathbf{ss}$
        \EndIf
    
    \EndWhile
\EndWhile
\end{algorithmic}
\end{algorithm}
Algorithm \ref{alg:GSO} implements the decimation procedure, whereas Algorithm \ref{alg:SA} is the simulated annealing. Notice that in the latter we have introduced a restarting criterion based on a moving energy threshold, that lowers any time we find a candidate pattern that is good enough, in terms of energy, or we do not find anything acceptable in 20 trials. The number of restarts needed increases exponentially with the size of the system, as in Fig. \ref{fig:restarts}, but the algorithm proved to be efficient up to sizes $N\simeq 1500$ for any noise level $\Delta$ we have tested, and above $N=2000$ when the noise is not too close to its critical value. 

\bibliography{main}

\end{document}